\documentclass{aa}
\usepackage{graphics}

\def\ltapprox{\raise 2pt \hbox {$<$} \kern-1.1em \lower 5pt \hbox {$\approx$}}
\def\ltsim{\raise 2pt \hbox {$<$} \kern-1.1em \lower 4pt \hbox {$\sim$}}
\def\gtsim{\raise 2pt \hbox {$>$} \kern-1.1em \lower 4pt \hbox {$\sim$}}

\begin{document}

\title{Particle acceleration in cooling flow clusters of galaxies: 
the case of Abell 2626}

\author{M. Gitti\inst{1,2,3}
\and G. Brunetti\inst{3}
\and L. Feretti\inst{3}
\and G. Setti\inst{2,3} 
}

\offprints{Myriam Gitti, c/o Istituto di Radioastronomia del CNR, via Gobetti
101, I--40129 Bologna, Italy; mgitti@ira.cnr.it}

\institute{Institut f\"ur Astrophysik, Universit\"at Innsbruck, 
Technikerstra{\ss}e 25, A-6020 Innsbruck, Austria
\and
Dipartimento di Astronomia, Universit\`a di Bologna,
via Ranzani 1, I--40127 Bologna, Italy
\and
Istituto di Radioastronomia del CNR,
via Gobetti 101, I--40129 Bologna, Italy 
}

\date{}

\abstract{
It has recently been proposed a theoretical model which accounts for the
origin of radio mini--halos observed in some cooling flow clusters 
as related to electron re--acceleration by MHD turbulence 
(Gitti, Brunetti \& Setti 2002).
The MHD turbulence is assumed to be frozen into the flow of the thermal ICM 
and thus amplified in the cooling flow region.
Here we present the application of this model to a new mini--halo candidate, 
the cluster A2626, and compare the results with those obtained for the
mini--halo in the Perseus cluster.
We present VLA data at 330 MHz and 1.5 GHz of the 
diffuse radio emission observed in A2626,
and we show that its main properties can be explained by the model. 
We find that the power necessary for the re--acceleration of the relic 
electron population is only a factor $\sim 0.7$ \% of
the maximum power that can be extracted by the cooling flow
(as estimated on the basis of the standard model).
We also discuss the observational properties of known mini--halos in 
connection with those of host clusters, showing that the 
radio power of mini--halos increases with the 
maximum power of cooling flows. 
This trend is expected in the framework of the model. 
Possible effects of new {\it Chandra} and {\it XMM--Newton} estimates of 
$\dot{M}$ on this trend are considered: we conclude that even if
earlier derived cooling rates were overestimated, cooling flow powers  
are still well above the radio powers emitted by mini--halos.
}

\authorrunning{M. Gitti et al.}
\titlerunning{Particle acceleration in cooling flow clusters of galaxies: 
the case of Abell 2626}                            

\maketitle


\section{Introduction}

A number of clusters of galaxies show extended synchrotron 
emission not directly associated with the galaxies but rather diffused into 
the intra--cluster medium (ICM). 
These radio sources have been classified in three classes: 
cluster--wide halos, relics and mini--halos (Feretti \& Giovannini 1996).
Both cluster--wide halos and relics have low surface brightness, large
size ($\sim$ 1 Mpc) and steep spectrum, but the former are located 
at the cluster centers and show low or negligible polarized emission, while
the latter are located at the cluster peripheries and are highly
polarized. 
They have been found in clusters which show significant evidence for an 
ongoing merger 
(e.g., Edge, Stewart \& Fabian 1992; Giovannini \& Feretti 2002). 
It was proposed that recent cluster mergers may play an 
important role in the re--acceleration of the radio--emitting relativistic 
particles, thus providing the energy to these extended sources
(e.g., Schlickeiser, Sievers \& Thiemann, 1987; Tribble 1993; 
Brunetti et al. 2001).
The merger picture is consistent with the occurrence of large--scale 
radio halos in clusters without a cooling flow, since the major merger 
event is expected to disrupt a cooling flow (e.g., Sarazin 2002 and references
therein).

In spite of the observed anti--correlation between the presence of
cooling flows and extended radio emission in clusters of galaxies,
there are several cooling flow clusters where the
relativistic particles can be traced out quite far
from the central galaxy, forming what is called a
``\textit{mini--halo}'' (e.g. Perseus: Burns et al. 1992;
Abell 2390: Bacchi et al. 2003).
Mini--halos are diffuse steep--spectrum radio sources, extended on a moderate 
scale (up to $\sim 500$ kpc), surrounding a dominant radio galaxy at the 
center of cooling flow clusters.
Until recently, because of the presence of the central radio galaxy, 
these sources have been considered of different nature
from that of extended halos and relics, and the problem of their origin
has never been investigated in detail.

Mini--halos do not appear as extended
lobes maintained by an Active Galactic Nucleus (AGN), as in classical radio
galaxies (Giovannini \& Feretti 2002), therefore their radio emission is
indicative of the presence of diffuse relativistic particles and magnetic
fields in the ICM.
Rizza et al.(2000) presented three-dimensional numerical 
simulations of perturbed jet propagating through a cooling flow atmosphere
in order to study the interaction between the radio plasma and the hot
ICM in cooling flow clusters containing steep--spectrum radio sources.
The evolution and spectrum 
of relativistic particles, however, is not 
considered in these simulations.
The point is that the radiative lifetime 
of the radio--emitting electrons injected at a given time in the
strong magnetic fields present in cooling flow regions is of the order of
$\sim 10^7$ - $10^8$ yr,
much shorter than the transport time necessary to cover $\sim$
hundred kpc scales, so that the diffuse radio emission from mini--halos
may suggest the presence of re--accelerated electrons.

More specifically, Gitti, Brunetti \& Setti (2002, hereafter GBS) suggested 
that the diffuse synchrotron emission from radio mini--halos is due to a relic 
population of relativistic electrons re--accelerated by MHD turbulence 
\textit{via} Fermi--like processes, with the necessary energetics 
supplied by the cooling flow. 

Alternatively, 
Pfrommer \& En{\ss}lin (2003) in 
a very recent paper
discussed the possibility that the radiating electrons
in radio mini--halos are of secondary origin and thus
injected during proton-proton collision in the ICM.

The main aim of the present work is the application of GBS's model to a new 
mini-halo candidate, A2626 ($z=0.0604$, Rizza et al. 2000), and the discussion 
of the observational properties of the population of radio mini--halos
so far discovered.

In Sect. 2 we briefly review GBS's model 
and its application to the Perseus  cluster. 
In Sect. 3 we consider the radio source observed in A2626: 
first, we present VLA data analysis and discuss the possibility
that this source belongs to the mini--halo class,
then we apply GBS's model to this cluster and discuss the results.
In Sect. 4 we present the observational properties of other radio mini--halo 
candidates in relation to those of host clusters and discuss them in the
framework of GBS's model.

For consistency with GBS, a Hubble constant
$\mbox{H}_0 = 50 \mbox{ km s}^{-1} \mbox{ Mpc}^{-1}$ 
is assumed in this paper, therefore at the distance of A2626  
$1'$ corresponds to $\sim$ 95 kpc.
The radio spectral index $\alpha$ is defined such as 
$S_{\nu} \propto \nu^{-\alpha}$ and,
where not specified, all the formulae are in cgs system.


\section{Electron re--acceleration in cooling flows}
\label{reacceleration.sec}

The radiative lifetime of an ensemble of relativistic electrons 
losing energy by synchrotron emission and inverse Compton (IC) scattering off 
the CMB photons is given by:
\begin{equation}
\label{tau_losses.eq}
\tau_{\rm sin+IC} \simeq \frac{2.5 \times 10^{13}}{\left[\left(
B/\mu\mbox{G}\right)^2 
+ \left(B_{\rm \tiny CMB}/\mu\mbox{G}\right)^2 \right] \, \gamma}
 \; \; \; \mbox{yr}
\end{equation}
where $B$ is the magnetic field intensity, $\gamma$ is the Lorentz 
factor and $B_{\rm \tiny CMB} = 3.18 (1+z)^2 \mu$G is the 
magnetic field equivalent to the CMB in terms of electron
radiative losses. 
In a cooling flow region the compression of the thermal
ICM is expected to produce a significant increase of the strength of
the frozen--in intra--cluster magnetic field and consequently
of the electron radiative losses.
Therefore, in the absence of a re--acceleration or continuous injection 
mechanisms,
relativistic electrons injected at a given time in these strong 
fields (of order of 10 - 20 $\mu$G, e.g., Ge \& Owen 1993; 
Carilli \& Taylor 2002) should already 
have lost most of their energy and the radio emission would not be observable 
for more than $\sim 10^7$ - $10^8$ yr.
This short lifetime contrasts with the diffuse radio emission observed
in mini--halos.

In order to evaluate the radiative losses in the cooling
flow region at any distance from the cluster center it is necessary 
to parameterize the radial dependence of the field strength, 
which depends on the compression of the thermal gas in the cooling
region (i.e., on $n(r/r_{\rm c})$, $r_{\rm c}$ being the
cooling radius).
However, it should be born in mind that while the X--ray
brightness and low resolution spectra are generally in agreement
with the standard cooling flow model, recent observations with the
Reflection Grating Spectrometer (RGS) on board {\it XMM--Newton}
do not show evidence for gas cooling at 
temperatures lower than 1-2 keV (e.g., Peterson et al. 2003) as
expected in the standard picture.
In addition, both {\it Chandra} and {\it XMM--Newton} results indicate
that the mass deposition rates in cooling flows have been previously
overestimated by a factor 5-10 (e.g., Fabian \& Allen 2003).
It is worth noticing that the new rates lead to mass 
values not too different from the large masses of cold gas 
derived from the studies of the CO emission line detected 
in a number of cooling flow cluster candidates
(Edge 2001; Salom\'e \& Combes 2003).
These findings, although not inconsistent with the idea
that the gas cools down and is thus compressed towards the central region,
point to a more complex situation than that described by the
standard cooling flow model.
Unfortunately, no successful model 
in alternative to the standard model
has been proposed yet and, therefore,
we will evaluate the
radial behaviour of the physical quantities in the ICM
by making use of the standard -- single phase -- cooling flow model.

In the framework of this model, the intensity 
of the frozen--in magnetic field increases as  
$B \propto r^{-2}$ for radial compression (Soker \& Sarazin 1990)
or $B \propto r^{-0.8}$ for isotropic compression (Tribble 1993).

\begin{figure}
\includegraphics{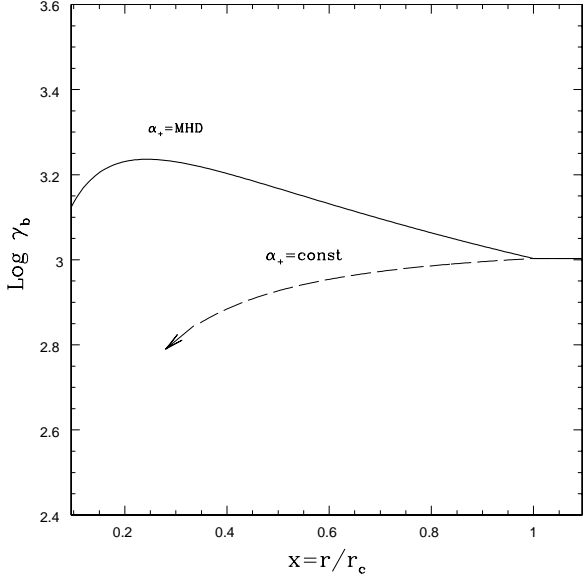}
\vspace{8cm}
\caption{Evolution of $\gamma_{\rm b}$ inside the cooling flow region 
for two different cases of re--acceleration: 
$\bullet \, \alpha_+$ given by Fermi--like processes related to the MHD 
turbulence in the cooling flow (bold line; 
Eq. \ref{gammabreak.eq} for typical values of the parameters), and
$\bullet \, \alpha_+$=constant (dashed line).}
\label{gammabreak.fig}
\end{figure}

The time evolution of the energy of a relativistic electron 
is determined by the competing processes of losses and re--accelerations 
(both related to the magnetic field) :
\begin{equation}
\dot{\gamma}(x) = - \beta(x) \gamma^2(x) + \alpha_+(x) \gamma(x) - \chi(x) 
\label{gammapunto.eq}
\end{equation}
where $x=r/r_{\rm c}$,
$\beta = \gamma^{-1}/\tau_{\rm sin+IC}$ is the coefficient of 
synchrotron and IC losses, $\alpha_+$ is the re--acceleration coefficient, 
which mimics the systematic rate of average energy increase 
resulting from stochastic acceleration (see GBS),
and $\chi$ the Coulomb loss term.
GBS developed a model for radio mini--halos consisting in the 
re--acceleration of relativistic electrons by MHD turbulence 
\textit{via} Fermi--like processes.
The MHD turbulence is assumed to be frozen into the flow of the thermal ICM 
and is thus amplified due to the compression of the turbulent coherence length
scale and the amplification of the magnetic field in the cooling flow
region.
In the present paper, we consider a coefficient for systematic 
re--acceleration given by (Melrose 1980):
\begin{equation} 
\label{coeffermi.eq}
\alpha_+(x) \approx (\pi/c) \, v_{\rm A}^2(x) \, l^{-1}(x) \,
[\delta B(x) / B(x)]^2 
\end{equation}
where $v_{\rm A} = \sqrt{B^2/4 \pi \rho}$ is the Alfv\'en velocity and $l$ 
is the characteristic MHD turbulence scale.
For simplicity we assume a fully developed turbulence with
$\delta B(x) / B(x) \sim {\rm const}$.
The energy at which the losses are balanced by the re--acceleration, 
 $\gamma_{\rm b}$,  
is obtained by Eq. \ref{gammapunto.eq} and, 
since Coulomb losses are nearly negligible
at such electron energies, it is 
$\gamma_b \approx \alpha_{+}/\beta$. 
The evolution of $\gamma_{\rm b}$ 
in the cooling flow region can be written in terms of the two free parameters 
$B_{\rm c}=B(r_{\rm c})$ and $l_{\rm c}=l(r_{\rm c})$, resulting
in (isotropic compression of the field is assumed,
as motivated in Sect. \ref{a2626_model.sec}):
\begin{equation}
\label{gammabreak.eq}
\gamma_{\rm b}(x) = \frac{1.23 \times 10^4 \, \left(\frac{B_{\rm c}}{1 \mu 
\mbox{\tiny G}}\right)^2 \, x^{-0.8} (\delta B_{\rm c}/B_{\rm c})^2 }
{\left(\frac{l_{\rm c}}{100 \mbox{ \tiny pc}}\right)
\left(\frac{n_{\rm c}}{10^{-3} \mbox{ \tiny cm}^{-3}} \right)
\left[ \left(\frac{B_{\rm c}}{1 \mu \mbox{\tiny G}} \right)^2  x^{-1.6}+
10 \right]}
\end{equation}
where $n_{\rm c}$ is the proton number density at $r_{\rm c}$,
and we used the relations $l(x) = l_{\rm c} \, x^{0.4}$ and 
$n(x) = n_{\rm c} \, x^{-1.2}$ (see GBS).
The behaviour of $\gamma_{\rm b}(x)$ is shown in Fig. \ref{gammabreak.fig}:
being able to make $\gamma_{\rm b}(x)$ weakly increasing inside the cooling
flow region, the re--acceleration due to MHD turbulence 
naturally balances the radial behaviour 
of the radiative losses.

Under these assumptions, the stationary spectrum of the relativistic 
electrons is given by: 
\begin{equation}
\label{ennegamma.eq}
N(\gamma, x) \approx N(\gamma_{\rm b})_{\rm c} 
\left({{\gamma_{{\rm b,c}} }\over
{\gamma_{\rm b}(x)}}\right)^2  {\rm e}^2 x^{-s} \cdot 
\left({{\gamma}\over{\gamma_{\rm b}(x)}}\right)^2 
{\rm e}^{(-2 \gamma/\gamma_{\rm b}(x))} 
\end{equation}
which is essentially peaked at $\gamma_{\rm b}$. 
Since the initial distribution of the number density and spectrum of the 
relativistic electrons, necessary to solve the spatial diffusion equation, 
are basically unknown, in obtaining Eq. \ref{ennegamma.eq} we parameterized 
the electron energy density as $\propto x^{-s}$, 
$s$ being a free parameter which will be constrained 
by model fitting (Sect. \ref{a2626_model.sec} and \ref{discussion.sec}).

\begin{figure*}
\includegraphics{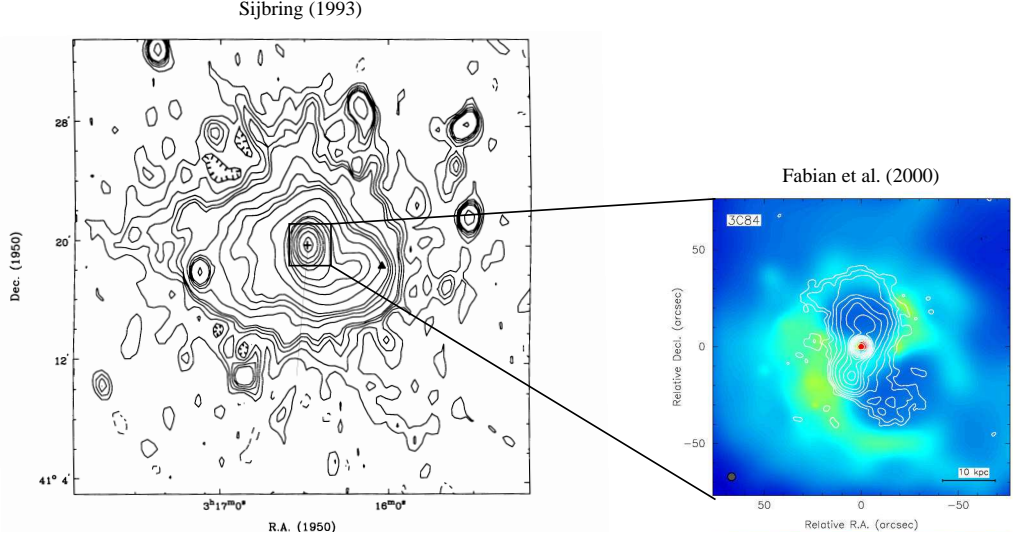}
\vspace{7.5cm}
\caption{\textbf{Left panel}: 327 MHz map of the mini-halo in the Perseus 
cluster at a resolution of $51'' \times 77''$ (Sijbring 1993). 
The cross indicates the position of the cD galaxy NGC 1275 
and the dashed line represents the direction considered 
for the application of GBS's model.
\textbf{~Right panel}: X--ray (grey scale)/radio (contours) overlay for the 
central part of the Perseus cluster around NCG 1275 (Fabian et al. 2000); 
X--ray data are obtained with {\it Chandra}. }
\label{perseus.fig}
\end{figure*}

The most striking evidence in favour of our model is provided by
the case of the Perseus cluster (A426, z=0.0183).
The diffuse radio emission from the mini--halo 
(see left panel in Fig. \ref{perseus.fig})
has a total extension of $\sim 15'$ (at the redshift of the cluster 
$1'$ corresponds to $\sim 30$ kpc) 
and its morphology is correlated with that of the cooling flow X-ray map
(Sijbring 1993; Ettori, Fabian \& White 1998).

On smaller scales ($\sim 1'$), there is evidence of interaction between 
the radio lobes of the central radiogalaxy 3C84 and the 
X--ray emitting intra--cluster gas (e.g., B\"ohringer et al. 1993; Fabian 
et al. 2000; see Fig. \ref{perseus.fig}, right panel).
A recent interpretation is that the holes in the 
X--ray emission are due to buoyant radio lobes 
which are currently expanding subsonically 
(Churazov et al. 2000; Fabian et al. 2002).  
It is important to notice that 
the spectral index in these lobes ranges from $\sim 0.7$
in the center to $\sim 1.5$ in the outer regions (Pedlar et al. 1990), 
which is a value similar to the spectrum of the mini--halo
extended over a scale $\sim 10$ times larger.  
Therefore, it is difficult to find a direct connection between the radio 
lobes and the large--scale mini--halo in terms of simple buoyancy or 
particle diffusion: 
the expansion and buoyancy of blobs would produce adiabatic
losses and a decrease of the magnetic field
causing a too strong steepening of the spectrum 
which would prevent the detection of large--scale radio emission.
In addition, the diffusion time 
($\propto R^2$, with $R=$ scale of interest; see also Sect. 
\ref{a2626_model.sec}) is about 100 times
longer than the radiative lifetime of the radio electrons.

Thus, if relativistic electrons are of
primary origin,
efficient re--acceleration mechanisms in the cooling flow 
region are necessary to explain the presence of the
large--scale radio emission in Fig. \ref{perseus.fig};
in particular, the detailed
modelling of the radio properties of the mini--halo 
in Perseus (brightness profile, 
integrated spectrum and radial spectral steepening) 
resulted in good agreement with the data in case
of isotropic compression of the magnetic field (GBS).


\section{Abell 2626: a new mini--halo candidate?}
\label{a2626.sec}

\begin{table*}
\begin{center}
\caption{VLA Data Archive}
\begin{tabular}{cccccccccc}
\hline
\hline
Proj Code & Source & Observation & Frequency & Bandwidth & Array & TOS & RA (J2000)& Dec (J2000)\\
~ & Name & Date & (MHz) & (kHz) & ~ & (h) & (h m s) & ($^{\circ}$ $'$ $''$) \\
\hline
~&~&~&~&~&~&~&~\\
AP001 & 3C464 & Oct-22-1985 & 307.50/327.50 & 3.125 & DnC & 1.5 & 23 36 30 & 21 08 44\\
AR279 & A2626 & May-04-1993 & 327.50/333.00 & 3.125 & B & 0.3 & 23 36 30 & 21 08 44\\
AR279 & A2626 & Jul-15-1993 & 1464.900 & 5.000 & C & 0.5 & 23 36 30 & 21 08 44\\
ROLA  & 4C2057 & Jul-20-1982 & 1464.900 & 5.000 & B & 0.7 & 23 36 30 & 21 08 06\\
\hline
\label{vladata.tab}
\end{tabular}
\end{center}
\end{table*}

The cluster A2626 hosts a relatively strong cooling flow (White, Jones \& 
Forman 1997) and contains an amorphous radio 
source near to the center (Roland et al. 1985; Burns 1990) which is
extended on a scale comparable to that of the cooling flow region with an
elongation coincident with the X--ray distribution (Rizza et al. 2000). 

X--ray image deprojection analysis of \textit{Einstein} IPC derives a
mass deposition rate 
$ \dot{M} \sim 53 \dot{\mbox{ M}}_{\odot} \mbox{ yr}^{-1}$,
a cooling radius $r_{\rm c} \sim 114$ kpc and an average 
temperature 
$kT \sim 3.1$ keV (White, Jones \& Forman 1997).
From Soker \& Sarazin (1990), one can estimate the proton number density 
at the cooling radius as:
\begin{equation}
\label{ennec.eq}
n_{\rm c} \approx 2.18 \times 10^{-3} \,\mbox{ cm}^{-3} 
\left(\frac{\dot{M}}{100 \mbox{ M}_{\odot} \mbox{ yr}^{-1}}
\right)  \left(\frac{r_{\rm c}}{\mbox{100 kpc}}\right)^{-3}
\end{equation}
so that 
$n_{\rm c} \sim 7.8 \times 10^{-4} \mbox{ cm}^{-3}$ with the values 
of $\dot{M}$ and $r_{\rm c}$ for A2626.


   \subsection{VLA archive data analysis}
   \label{vla.sec}

In order to extend the application of GBS's model to A2626, 
we have requested and analyzed some of the VLA archive data 
(Table \ref{vladata.tab}) of A2626 with the aim to derive the surface 
brightness map, the total spectral index and the spectral index distribution 
of the diffuse radio emission.

\begin{figure}
\includegraphics{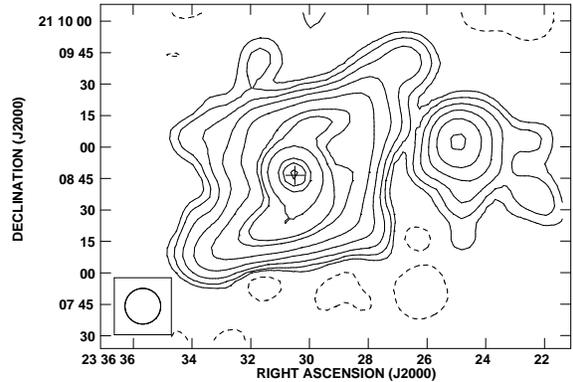}
\vspace{6cm}
\caption{
1.5 GHz VLA map of A2626 at a resolution of $17'' \times 17''$. 
The contour levels are $-0.06$ (dashed), 0.06, 0.12, 0.24, 0.48, 0.96, 1.92,
2.5, 4, 8, 10, 12 mJy/beam. The r.m.s. noise is 0.02 mJy/beam.
The cross indicates the position of the cluster center.}
\label{mappa1.5.fig}
\end{figure}

\begin{figure}
\includegraphics{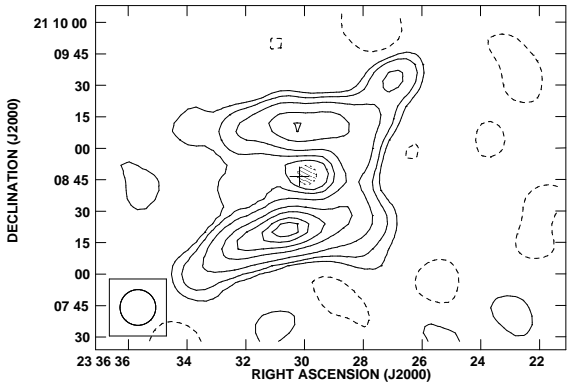}
\vspace{6cm}
\caption{
330 MHz VLA map of A2626 at a resolution of $17'' \times 17''$. 
The contour levels are $-8.5$ (dashed), 8.5, 17, 34, 68, 100, 135,
160 mJy/beam. The r.m.s. noise is 3.1 mJy/beam. The cross indicates the 
position of the cluster center and the dashed region defines the 
central brightness hole.}
\label{mappa330.fig}
\end{figure}

Standard data reduction was done using the National Radio Astronomy Observatory
(NRAO) AIPS package.
We used the 1.5 GHz C array data and the 330 MHz B+DnC array data
to produce low resolution images with a circular restoring beam of 
17 arcsec. The imaging procedure at each frequency was performed using 
data with matched uv coverage. 
These images (Fig. \ref{mappa1.5.fig} and \ref{mappa330.fig})
allow to derive morphological and spectral information of the 
diffuse emission.
The 1.5 GHz map (Fig. \ref{mappa1.5.fig}) shows an unresolved core and 
a diffuse boxy--shaped emission extended for $\sim 2'$, 
corresponding to about 190 kpc.
No significant polarized flux is detected, leading to a 
polarization upper limit
of $<2$ \%. 
An unrelated source is present to the 
west of the diffuse emission, 
with a total flux density of $\sim$3.9 mJy.

The diffuse structure 
at 330 MHz (Fig. \ref{mappa330.fig})
is smaller in extent,
because of the lower sensitivity. It consists of
two elongated almost parallel features located
to the north and south of the core, respectively.
No radio emission is detected at the location of the
1.5 GHz radio nucleus. Assuming for the 330 MHz core flux
an upper limit of 3$\sigma$, we obtain that 
the radio core  has an inverted spectrum (Table \ref{risradio.tab}).
The total flux density of the diffuse emission is 
$S_{330} \sim 1$ Jy. 
The unrelated source, detected at 1.5 GHz to the
west of the diffuse
radio emission, is not detected at 330 MHz; this implies
a spectrum with $\alpha^{1.5}_{0.3}$ \ltsim 0.6.

\begin{figure}
\includegraphics{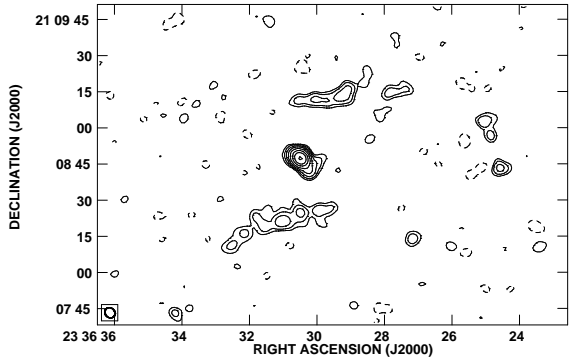}
\vspace{6cm}
\caption{
1.5 GHz VLA map of A2626 at a resolution of $4.5'' \times 3.9''$. 
The contour levels are $-0.09$ (dashed), 0.09, 0.18, 0.36, 0.7, 1.4, 3,
5.5, 11, 20 mJy/beam. The r.m.s. noise is 0.03 mJy/beam}
\label{4c2057.fig}
\end{figure}

In order to estimate the total integrated flux density
of the diffuse radio emission at 1.5 GHz and derive the surface 
brightness map and spectral trend, it is necessary 
to subtract the emission from the central nuclear source.
One way is to make a high--resolution image and then
extract the clean components of the central source.
By using the 1.5 GHz B array data 
we produced the high resolution
image (Fig. \ref{4c2057.fig}) with a restoring beam of $4.5 \times 3.9$ arcsec.
The central component is found to consist of an unresolved
core, plus a jet-like feature pointing to the south-western direction.
The extended emission is resolved out and two elongated parallel features of 
similar brightness and extent are detected. 
The flux density of the central component, the short jet included,
is $\simeq 13.5$ mJy, in agreement with Roland et al. (1985).
The clean components of the central discrete source (jet included) 
were then restored with a circular beam of 17 arcsec and subtracted from the 
low--resolution map.
This subtraction allows to derive a good estimate of the 
total flux density of the diffuse emission ($S_{1.5} \sim 29$ mJy).

\begin{figure}
\includegraphics{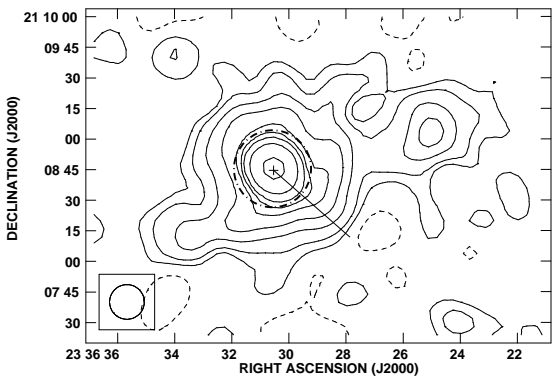}
\vspace{6cm}
\caption{
1.5 GHz VLA map of A2626 at a resolution of $17'' \times 17''$
after the subtraction of the two elongated features visible in
Fig. \ref{4c2057.fig}. The cross indicates the position of the cluster
center. 
The contour levels are $-0.18$ (dashed), 0.18, 0.36, 0.72, 1.2, 1.85, 2.3
mJy/beam. The r.m.s. noise is 0.08 mJy/beam.
The dash--dotted circle defines the nuclear region excluded in modelling the 
diffuse radio emission,
while the solid line represents the direction we have considered for the
surface brightness profile (see Fig. \ref{fit.fig}).
}
\label{a2626_minihalo.fig}
\end{figure}

As discussed in Sect. \ref{a2626_model.sec} and \ref{discussion.sec}, 
we believe that the elongated structures visible in Fig. \ref{4c2057.fig}
are distinct 
(or that they may represent an earlier evolutionary stage)
from the diffuse emission, and that the diffuse radio source
may belong to the mini--halo class.
The total flux density of these structures is $\sim 6.6$ mJy, 
thus contributing to only $\sim$ 20 \% of the flux of 
the diffuse radio emission.
Since we are interested in studying and modelling the diffuse radio emission,
we produced a new low--resolution map at 1.5 GHz where these discrete
radio features have been subtracted (Fig. \ref{a2626_minihalo.fig}).
The subtraction procedure was similar to that used for the subtraction
of the central discrete source.
Of course the resulting map (Fig. \ref{a2626_minihalo.fig}) has a r.m.s.
noise much higher than that of original map (Fig. \ref{mappa1.5.fig}).
We notice that after the subtraction of the elongated features
the morphology of the diffuse radio emission becomes roughly circular,
thus warranting the application
of a spherical model in Sect. \ref{a2626_model.sec}. 
To be conservative, the region in which the surface brightness 
is affected by the central emission (within the dash--dotted circle 
in Fig. \ref{a2626_minihalo.fig}) has been excluded in modelling the diffuse 
radio emission (see Sect. \ref{a2626_model.sec}).

Unfortunately, due to the lack of a high--resolution image,
the two elongated features can not be subtracted at 330 MHz as well.
Therefore, in deriving the spectral information 
of the diffuse emission we considered for consistency
the low--resolution images in Fig. \ref{mappa1.5.fig} 
and \ref{mappa330.fig}, which both include the 
contribution of the two features to the total flux.

The integrated spectral index of the diffuse emission between 
$\nu=330$ MHz and $\nu = 1.5$ GHz is $\alpha \sim 2.4$. 
If the two features consist of relatively fresh
injected plasma (as discussed in Sect. \ref{discussion.sec}),
their spectrum is expected to be relatively flat
and thus the real spectrum of the diffuse emission 
would be slightly steeper
($\alpha \sim 2.5$ in the extreme case in which the two 
features do not contribute
to the flux measured at 330 MHz).

\begin{figure}
\includegraphics{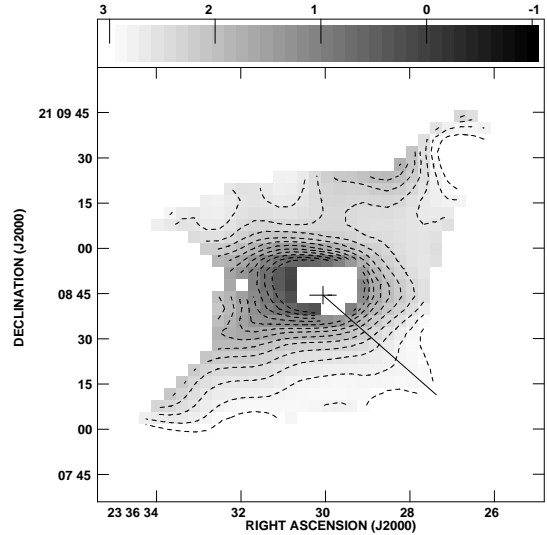}
\vspace{7cm}
\caption{Spectral index distribution between $\nu=330$ MHz and $\nu = 1.5$ GHz
at a resolution of $17'' \times 17''$. The contours are in steps of 0.2, 
where the lightest grey is in the range 2.8 to 3.
The lighter the grey, the steeper the spectral index.
We have excluded the region in which the error is $>0.2$.
The cross indicates the position of the subtracted radio core and the solid
line represents the direction we have considered for the spectral steepening
profile (see Fig. \ref{irr.fig}).}
\label{mappaindice.fig}
\end{figure}

In Fig. \ref{mappaindice.fig} we show a grey scale image
of the spectral index map of the diffuse radio emission between  
$\nu=330$ MHz and $\nu = 1.5$ GHz.
The spectrum steepens from the central region towards the north and south
direction, with the spectral index increasing from $\sim$1.2
to $\sim$3. 
The steepening in the northern region is slightly smoother than in the 
southern one.

Electrons emitting at frequencies $\sim$ 1 GHz in magnetic fields of the 
order of $\sim$ 1-3 $\mu$G (see Sect. \ref{discussion.sec}) have a
radiative lifetime (Eq. \ref{tau_losses.eq}) $\sim 7 \times 10^7$ yr.

The radio results are summarized in Table \ref{risradio.tab}.

\begin{table}
\begin{center}
\caption{Radio results for A2626}
\begin{tabular}{ccccc}
\hline
\hline
Emission & $S_{330}$ & $S_{1.5}$ & Size & $\alpha$ \\
~&  (mJy) & (mJy) & (arcsec$^2$) & ($S_{\nu} \propto \nu^{-\alpha}$)\\   
\hline
~&~&~&~\\
Nuclear & $<$9.3 & $13.5 \pm 1.5$ & -  & \ltsim $-0.25$\\
Diffuse & $990 \pm 50$ & $29 \pm 2$ & $135 \times 128$ & $2.37 \pm 0.05$ \\
\hline
\label{risradio.tab}
\end{tabular}
\end{center}
\end{table}


   \subsection{Model for A2626}
   \label{a2626_model.sec}

As presented in Sect. \ref{vla.sec}, the extended radio 
source observed at the center of A2626 is characterized by amorphous 
morphology, lack of polarized flux and very steep spectrum which
steepens with distance from the center.  
Finally, the morphology of the diffuse radio emission
is similar to that of the cooling flow region (Rizza et al. 2000). 
These results indicate that the diffuse radio source may be 
classified as a mini--halo.

In addition, 
we notice that there are some concerns in interpreting this source 
without assuming the presence of particle re--acceleration.
One possibility would be that the radio--emitting region is being 
supplied with fresh relativistic plasma that ultimately comes from 
the nucleus. 
Thus the observed diffuse radio emission would 
result from a bubble--like structures   
expanding into the surrounding cluster gas, as suggested 
for the large--scale radio structure of M87 
presented by Owen, Eilek \& Kassim (2000). 
Following the model adopted by these authors for M87,
by assuming that the bubble pressure remains comparable to the external 
pressure (that is approximately constant in a cooling flow region, 
resulting 
$\sim 7.7 \times 10^{-12} \mbox{ dyn cm}^{-2}$ at $r_{\rm c}$)
we estimate that the age of the radio structure of A2626 
at a radius $\sim$ 100 kpc would be
$\sim 1.2 \times 10^9 \, P_{44}^{-1}$ yr
(we adopted an internal adiabatic index of  
$\Gamma = 4/3$ and 
the jet power $P$ in units of $10^{44}
\mbox{ erg s}^{-1}$).
We notice that, with a typical value $P_{44} \sim 1$, 
this age would be 
at least 10 times longer than the lifetime of the radio--emitting 
electrons, thus in situ acceleration would be necessary 
to energize 
the radio electrons. 
In addition we notice that 
this scenario is not able to explain the observed spectral 
steepening of the radio emission with distance.
Another possibility, which instead may be 
able to explain a radial 
spectral steepening, would be particle diffusion out of the central region.
The diffusion length is defined as $R_{\rm diff} \simeq \sqrt{6 {\cal{K}} 
\tau_{\rm diff}}$ (e.g., Fujita \& Sarazin 2001), 
where $\cal{K}$ is the spatial diffusion coefficient
and $\tau_{\rm diff}$ the diffusion time.
By assuming the commonly adopted Kolmogorov spectrum of the 
magnetic field
it is possible to verify that for radio--emitting electrons 
(which typically have
Lorentz factors $\gamma \sim 10^4$ and radiative
lifetimes $\sim 10^8$yr) the diffusion length 
(by adopting the parallel diffusion coefficient, 
$\cal{K}_{\Vert}$)
during their lifetime is $R_{\rm diff}$ \ltsim
30 kpc (e.g., Brunetti 2003). 
This diffusion length is thus
a factor 3 shorter than the scale of interest in 
the radio source of A2626 ($\sim 100$ kpc).
As a consequence, diffusion may be an efficient process 
only for electrons with energy  
10 times smaller than that of
those responsible for the large--scale radio 
emission. 
Anomalous diffusion 
may considerably increase the propagation of electrons
perpendicular to the magnetic field lines
(e.g., Duffy et al. 1995).
This kind of diffusion increases with increasing the turbulence
energy density; however, the ratio between anomalous 
and parallel diffusion coefficients is (Giacalone \& Jokipii 1999;
En{\ss}lin 2003) 
${\cal{K}}_{\rm a}/{\cal{K}}_{\Vert} \sim 0.04 (\delta B/B)^4$
and thus parallel diffusion still remains the most efficient way for
particle diffusion also in the case of relatively
powerful turbulence (i.e., $\delta B \sim B$).
One possibility to enhance anomalous diffusion is to 
significantly increase the turbulence energy density
and change, at the same time,
the ratio between large--scale and small--scale turbulence.
This possibility has been discussed in the literature 
to allow a more efficient escape of cosmic rays from 
the cocoon of radio sources (En{\ss}lin 2003).
However, given typical conditions in a cooling flow region,
an unlikely high level of MHD turbulence should be injected 
(e.g., $\delta B / B \sim$ 15-20)
in order to guarantee an electron diffusion length of 
$\sim 100$ kpc in $< 10^8$ yr via enhanced anomalous diffusion, 
but this would also yield an extremely efficient particle acceleration
via wave-particle scattering.

Given the above difficulties in explaining the amorphous radio structure
observed in A2626, we apply GBS's model.
However, the application is not straightforward as in the case of Perseus
because of the presence of the two structured radio features observed 
about $20''$ on either side of the nucleus in the high resolution image 
(Fig. \ref{4c2057.fig}), which may indicate that there is still some 
injection of relatively young ($\sim 10^8$ yr) relativistic plasma in the 
cooling flow region. 
This suggests that A2626 has physical properties in between 
those of the well known case of M87 (in which there is only
marginal evidence for electron re--acceleration, e.g., Owen, Eilek \& Kassim 
2000) and of the prototype of mini--halo, in the Perseus cluster.

One important conclusion reached in the application of the 
GBS's model to the Perseus
cluster is that the results are compatible with the observations
by assuming an isotropic compression of the magnetic field
in the cooling flow region.
This is consistent with the fact that
the turbulent velocity results greater than the
mean inflow velocity at the cooling radius,
$v_{\rm F,c} \sim 21 \left(\dot{M}/100 
\mbox{ M}_{\odot} \mbox{ yr}^{-1}\right)
 \left(r_{\rm c}/\mbox{100 kpc}\right)^{-2} 
\left(n_{\rm c}/10^{-3} \mbox{ cm}^{-3}\right)^{-1}\\ 
\mbox{km s}^{-1}$ (Soker \& Sarazin 1990), 
and isotropic compression of the 
field is expected (Tribble 1993).
By assuming a relatively powerful
turbulence (i.e.,
$\delta B \sim B$), the turbulence energy density 
at the cooling radius is 
$\eta \, n_{\rm c} m_{\rm p} v_{\rm T,c}^2 \sim B_{\rm c}^2/8\pi$
($\eta < 1$ is a fudge factor) and thus the 
velocity of the eddies of turbulence would be 
$
v_{\rm T,c} \sim 90 \, (B_{\rm c}/1.6 \mu G) \,
\eta^{-1/2} \, (n_{\rm c}/ 8 \times 10^{-4} \mbox{cm}^{-3})^{-1/2} 
\mbox{ km s}^{-1}$.
This value is greater than that of the mean inflow velocity
of A2626 ($v_{\rm F,c} \sim 10 \mbox{ km s}^{-1}$) and 
thus we considered only the case of an isotropic compression of the 
magnetic field.

\begin{figure}
\includegraphics{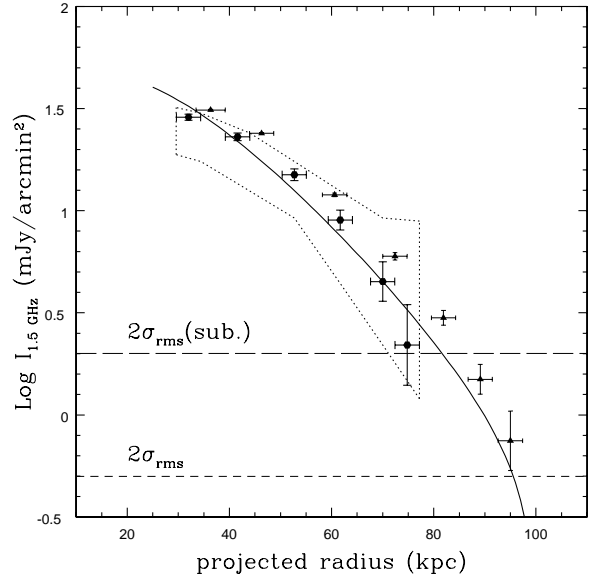}
\vspace{8cm}
\caption{
Fit to the surface brightness profile (black circles) observed at 1.5 GHz
along the radial direction indicated in Fig. \ref{a2626_minihalo.fig}.
The dotted line defines the observed region of 
values of the brightness distribution at any radius, 
while the large dashed line represents the 
2 r.m.s. noise level of the map subtracted from the elongated radio features
(the errorbars are at 1$\sigma$).
The fit is obtained with the following set of parameters: 
$B_{\rm c}=1.5 \mu\mbox{G}$, 
$l_{\rm c}=175 \mbox{ pc}$, $s=-0.5$. 
For comparison, we also report (with triangles) the brightness profile taken 
along the same radial direction 
from the 1.5 GHz map including the contribution of the 
elongated features (Fig. \ref{mappa1.5.fig}). 
The small dashed line represents the 2 r.m.s. noise level of the 
non--subtracted map.
}
\label{fit.fig}
\end{figure}

We have already noticed 
that after the subtraction of the two elongated features,
the morphology of the diffuse radio emission becomes roughly spherical
(see Fig. \ref{a2626_minihalo.fig}),
thus justifying the application to A2626 of GBS's model, which indeed assumes
spherical symmetry. In particular, in fitting the brightness profile
we are authorized to choose a particular radial direction and give
the deviations from the spherical symmetry in other directions with respect
to the one considered.

In order to test the prediction of the model 
 we have calculated the following expected observable properties:

\textit{total spectrum}:
the total synchrotron spectrum is obtained by integrating the synchrotron 
emissivity from an electron population with the energy
distribution given by Eq. \ref{ennegamma.eq} over the cluster volume; 

\textit{brightness profile}: 
the surface brightness profile expected by the model is obtained
by integrating the synchrotron emissivity at 1.5 GHz along the line of sight;

\textit{radial spectral steepening}:
at varying projected radius, we obtain the surface brightness at 330 MHz and 
1.5 GHz, and compute the spectral index between these two frequencies.\\
The expected brightness profile and radial spectral steepening 
in the model are compared to the observed profiles along the radial 
direction indicated in Fig. \ref{a2626_minihalo.fig} and 
\ref{mappaindice.fig}. 
Since it is not possible to subtract the elongated features
at 330 MHz (see Sect. \ref{vla.sec}), their contribution 
to the total spectrum is accounted for by the
errorbars.

\begin{figure}
\includegraphics{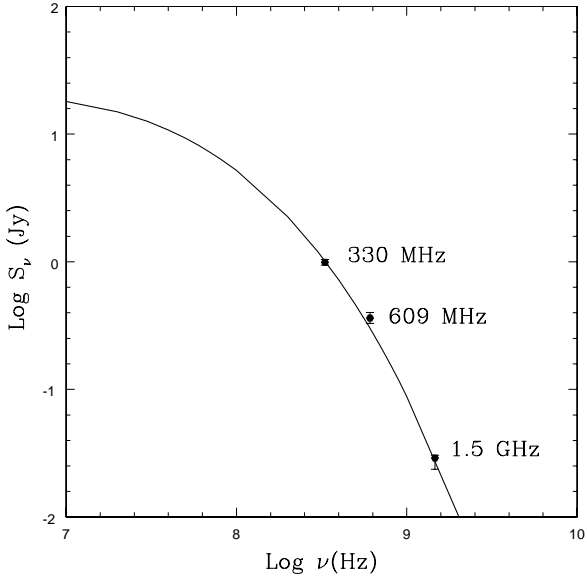}
\vspace{8cm}
\caption{
Fit to the total spectrum of the synchrotron emission obtained with the same 
set of parameters of Fig. \ref{fit.fig}. 
The flux densities observed at 330 MHz and 1.5 GHz are taken from Table 
\ref{risradio.tab}, while the flux density at 609 MHz is obtained by 
subtracting the estimated core emission from the total flux given by 
Roland et al. (1985).
The contribution of the elongated features to the flux of the diffuse 
emission (as estimated in Sect. \ref{vla.sec}) is included in the errorbars.
}
\label{flussi.fig}
\end{figure}

\begin{figure}
\includegraphics{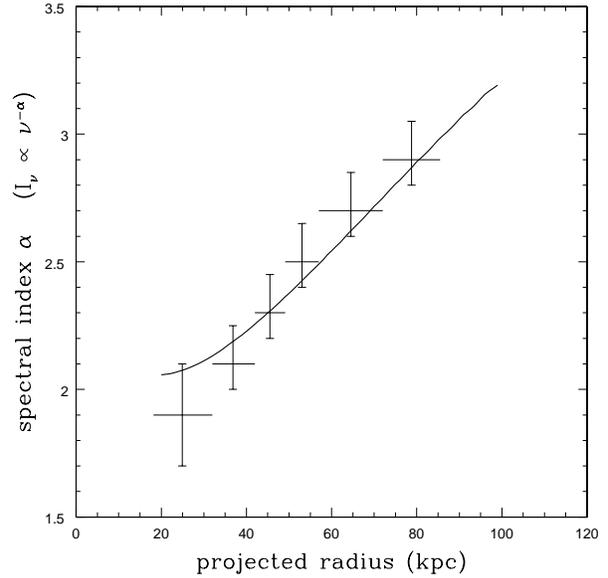}
\vspace{8cm}
\caption{Fit to the radial spectral steepening between $\nu=330$ MHz and 
$\nu = 1.5$ GHz obtained with the same set of parameters of Fig. 
\ref{fit.fig}. 
The data are taken from the spectral index distribution along the 
direction indicated in Fig. \ref{mappaindice.fig}.
The effect induced by the contribution of the elongated features
to the flux of the diffuse emission
(as estimated in Sect. \ref{vla.sec}) is included in the errorbars.
}
\label{irr.fig}
\end{figure}

The values of the parameters required by the model to match the three 
observational constraints 
at the 90\% confidence level are:
$-1 < s < 0$, $B_{\rm c} \sim 1.2$ - $1.6 \, \mu$G, 
$l_{\rm c}/(\delta B_{\rm c}/B_{\rm c})^2 \sim 120$ - $180$ pc, 
where the lower $s$ corresponds to the lower $B_{\rm c}$ 
and $l_{\rm c}/(\delta B_{\rm c}/B_{\rm c})^2$.
For these parameters, one obtains that the break
energy at the cooling radius is $\gamma_{\rm b,c} \simeq 1100$.
In Fig. \ref{fit.fig}, \ref{flussi.fig} and \ref{irr.fig} we show the fits 
to the surface brightness profile, total spectrum and 
radial spectral steepening for one set of the parameters which best 
reproduces all the observational constraints.

   \subsection{Discussion}
   \label{discussion.sec}

The physical implications derived by applying GBS's
model to A2626 are discussed 
and compared with the results obtained for the Perseus 
cluster. 
For completeness the X--ray and radio data for these clusters are 
listed in Table \ref{confr_dati.tab} while the model results 
for both clusters are summarised in Table \ref{confr_ris.tab}.

It is worth pointing out that GBS's model is able to reproduce all the
observational constraints of A2626 for physically--meaningful 
values of the 
parameters (Table \ref{confr_ris.tab}). 
First of all we found that in the case of A2626  
the range of values obtained for $B_{\rm c}$,
although somewhat higher than that of Perseus,
is in agreement with the measurements of magnetic field strengths in the ICM
(Carilli \& Taylor 2002 and references therein). 

We also notice that the
$l_{\rm c}/[\delta B_{\rm c} / B_{\rm c}]^2$ value
required by the
model for A2626 is higher than that found for Perseus (see column 3 of
Table \ref{confr_ris.tab}). 
Even though we can not discriminate between the two contributions to this
parameter, it is more likely that $[\delta B_{\rm c} / B_{\rm c}]^2$ 
in A2626 is smaller than in Perseus, as $l_{\rm c}$ depends on the 
micro--physics and is not expected to change considerably. 
In particular, in the general theory of turbulent plasma one can 
calculate the wavelength which carries most of the turbulent energy 
in a spectrum of Alfv\'en waves (e.g. Tsytovich 1972).
When applied to the case of the ICM, with standard values of the physical
parameters, it gives results of the order of tens to hundreds
pc (GBS).

\begin{table*}
\begin{center}
\caption{Observational data for the cooling flow and mini--halo in Perseus 
and A2626}
\begin{tabular}{cc|c|c|c|c||c|c|c|}
\multicolumn{1}{c}{~} & \multicolumn{5}{c}{X--RAY DATA} & \multicolumn{3}{c}{RADIO DATA}\\ 
\hline
\hline
\multicolumn{1}{|c||}{Cluster} & $\dot{M}$ & $r_{\rm c}$ & $kT$ & $n_{\rm c}$ & $P_{\rm CF}$ & $P_{1.5}$ & $r_{\rm mh}$& $\alpha$\\
\multicolumn{1}{|c||}{~} & ($\mbox{M}_{\odot} \mbox{ yr}^{-1}$) & (kpc) & (keV) & 
($\mbox{cm}^{-3}$) & (erg s$^{-1}$) & (W Hz$^{-1}$) &(kpc)& ($S_{\nu} \propto \nu^{-\alpha}$)\\
\hline
\multicolumn{1}{|c||}{~}&~&~&~&~&~&~&~&~\\
\multicolumn{1}{|c||}{Perseus} & $519^{+3}_{-17}$ & $210^{+100}_{-20}$ & $6.33^{+0.21}_{-0.18}$ & $1.2 \times 10^{-3}$ &$2.6 \times 10^{44}$ & $4.4 \times 10^{24}$ & $\sim 300$ & $\sim 1.2$\\
\multicolumn{1}{|c||}{A2626} & $ 53^{+36}_{-30}$ & $114^{+50}_{-59}$ & $3.1^{+0.5}_{-1}$ & $7.8 \times 10^{-4}$ & $1.2\times10^{43}$ & $4.3 \times 10^{23}$ & $\sim 100$ & $\sim2.4$\\  
\hline
\label{confr_dati.tab}
\end{tabular}
\end{center}
\textit{Notes}: 
Columns 2 and 3 list the cooling flow parameters, taken from Ettori, Fabian 
\& White (1998) with \textit{ROSAT} PSPC for Perseus and from 
White, Jones \& Forman (1997) with \textit{Einstein} IPC for A2626.
Column 4 lists the average temperature of the ICM, taken from Allen et al.
(1992) with \textit{Ginga} for Perseus and from White, Jones \& Forman (1997) 
with \textit{Einstein} IPC for A2626. 
Column 5 lists the electron number density $n_{\rm c}$, estimated 
from Eq. \ref{ennec.eq}.
Column 6 lists the power supplied by the cooling flow as estimated from
X--ray data by Eq. \ref{pcf.eq}.
Columns 7, 8 and 9 list the physical properties of the mini--halo: total 
power at 1.5 GHz, radius, and integrated spectral index between
$\nu \sim 330$ MHz and $\nu \sim 1.5$ GHz. 
Radio data for Perseus are taken from Sijbring (1993).
\end{table*}

\begin{table*}
\begin{center}
\caption{Model results for Perseus and A2626}
\begin{tabular}{cc|c|c||c|c|c|c|c|c|}
\multicolumn{1}{c}{~} & \multicolumn{3}{c}{MODEL PARAMETERS} & \multicolumn{6}{c}{MODEL RESULTS}\\
\hline
\hline
\multicolumn{1}{|c||}{Cluster} & $B_{\rm c}$ & $l_{\rm c}/[\delta B_{\rm c} / B_{\rm c}]^2$ & s & $\gamma_{\rm b,c}$ & $N(\gamma_{\rm b})_{\rm c} \cdot \gamma_{\rm b,c}$ & $E_{\rm e}$ & $ N_{\rm e}$ & $P_{\rm e}$ & $\epsilon$\\
\multicolumn{1}{|c||}{~} & ($\mu \mbox{G}$) & (pc) &~ & ~ & ($\mbox{cm}^{-3}$)& (erg) &~ & (erg s$^{-1}$) & (\%)\\
\hline
\multicolumn{1}{|c||}{~}&~&~&~&~&~&~&~&~&~\\
\multicolumn{1}{|c||}{Perseus} & 0.9 - 1.2 &  35 - 60  & $\sim 2 $ & 1600 & $ 2.2 \times 10^{-12}$ & $1.6 \times 10^{58} $ & $1.6 \times 10^{61} $ & $ 8.9 \times 10^{41}$ & $ 0.34$\\
\multicolumn{1}{|c||}{A2626} & 1.2 - 1.6 & 120 - 180  & $\sim -0.5 $ & 1100 & $ 8.5 \times 10^{-12}$ & $2.2 \times 10^{57} $ & $3.2 \times 10^{60} $ & $ 8.3 \times 10^{40}$ & $ 0.69$\\  
\hline
\label{confr_ris.tab}
\end{tabular}
\end{center}
\textit{Notes}: 
Columns 2, 3 and 4 list the parameters of the model derived by fitting all the
observational constraints. Column 5, 6, 7, 8 and 9 list the physical 
properties derived by the model: respectively, the break energy of the 
electron spectrum at $r_{\rm c}$, the number density of electrons with energy
$\gamma_{\rm b} m_{\rm e} c^2$ at $r_{\rm c}$, the energetics associated 
with the electrons re--accelerated in the cooling flow region 
(Eq. \ref{energetica.eq}), their total number and the power necessary to 
re--accelerate them (Eq. \ref{power_el.eq}).
Column 10 lists the efficiency $\epsilon$ required by the cooling flow for 
re--accelerating the relativistic electrons, given in percentage of 
$P_{\rm CF}$ taken from column 6 of Table \ref{confr_dati.tab}. 
The results for Perseus are taken from GBS and modified according to Eq.
\ref{coeffermi.eq} and \ref{ennegamma.eq}. 
See text for details.
\end{table*}

Concerning the energy density of the relativistic electrons, we find that 
it is approximately constant inside the cooling flow region ($s \sim 0$).
Since $\gamma_{\rm b}$ depends very weakly on the radial distance, 
this means that the radial distribution of the number density of the 
electrons before the re--acceleration is nearly constant, producing a
sort of ``plateau'' of relic electrons in the cooling flow region.
On the other hand, in the case of Perseus, GBS 
found that the energy density of the relativistic electrons 
increases towards the center.
A possible qualitative explanation of the ``plateau'' 
is that past radio activity may have released electrons
in the cooling flow region.
That this might be the case is suggested by the presence of
the two extended features in Fig.~5 which, indeed,  
may be buoyant plumes recently
ejected by the central source.
As shown by several authors 
(e.g., Gull \& Northover 1973; Churazov et al. 2000; 
Br\"uggen \& Kaiser 2002), the outflow 
is accompanied by adiabatic expansion and further mixing
of the energetic relativistic electrons with the ambient ICM;
the time--scale of this process ($\sim 10^7$ - $10^8$ yr) is 
comparable to (or bigger than)
the lifetime of the electrons producing synchrotron radiation. 
The disruption of the bubbles produced in past radio
outbursts would then have left a population of 
relic relativistic electrons mixed with the thermal plasma 
in the cooling flow region.
These relic electrons could diffuse in the thermal plasma up to
$\sim 100$ kpc scale in a few Gyr (e.g., Brunetti 2003) 
filling the whole region 
inside the two elongated parallel features observed in 
Fig. \ref{4c2057.fig}, thus forming the ``plateau'' of relic
electrons requested by our modelling.

Note that we found an electron distribution with 
$\gamma_{\rm b} \sim 10^3$.
We stress that this is not inconsistent with the value 
$\gamma \sim 10^4$ adopted in the discussion of diffusion of 
radio--emitting electrons (Sect. \ref{a2626_model.sec}), 
because a re--accelerated electron 
population with $\gamma_{\rm b} \sim 10^3$ is able to emit at
GHz frequencies thanks to the shape of the re--accelerated spectrum 
which is peaked at $\sim \gamma_b$
(see Eq. \ref{ennegamma.eq}), while 
the spectrum of non re--accelerated electrons 
(a power--law with a high energy cut--off at such a $\gamma_{\rm b}$)
does not allow to emit efficiently in this band. 
As a consequence, without acceleration (i.e., simple
diffusion model), electrons with $\gamma_{\rm b} \sim 10^4$
are necessary to account for the radio flux.

The energetics associated with the population of electrons re--accelerated
in the cooling flow region can be estimated as:

\begin{equation}
\label{energetica.eq}
E_{\rm e} \approx 23 \frac{r_{\rm c}^3 \, m_{\rm e} c^2 
N(\gamma_{\rm b})_{\rm c} \,  \gamma_{\rm b,c}^2}{(3-s)} \; \; \; \mbox{erg}
\end{equation}
where $N(\gamma_{\rm b})_{\rm c}$ is the number density (per $\gamma$ unit) 
of electrons with energy $\gamma_{\rm b} m_{\rm e} c^2$ at $r_{\rm c}$.

The total number of relativistic electrons can be estimated from 
the energetics as:
$N_{\rm e} \sim 4 E_{\rm e}/(3 m_{\rm e} c^2 \, \gamma_{\rm b,c})$.

With the model parameters found for A2626, one obtains 
$ E_{\rm e} \sim 2.2 \times 10^{57} $ erg and 
$N_{\rm e} \sim 3.2 \times 10^{60}$.
It is worth noticing that both the energetics and the number of 
electrons re--accelerated in the cooling flow region are 
about one order of magnitude smaller than those found in Perseus.
This is consistent with the fact that the radio power of the mini--halo 
in A2626 is about one order of magnitude smaller than that in Perseus.

The power $ P_{\rm e}$ 
necessary to re--accelerate the emitting electrons is given by
the minimum energy one must supply to 
balance the radiative losses of these electrons:
$P_{\rm e} \approx m_{\rm e} c^2 \beta \, 
\gamma_{\rm b,c}^2 \cdot  N_{\rm e}$,
where $\beta$ is the same as in Eq. \ref{gammapunto.eq}.
By assuming an average magnetic field in the cooling flow region
of order $\sim 3 \mu$G 
(this value is justified by considering 
the intensity
$B_{\rm c}$ obtained in the model
and the radial behaviour of field amplification expected in the case of 
isotropic compression), we obtain:

\begin{equation}
P_{\rm e} \approx 2 \times 10^{-26} \, \gamma_{\rm b,c}^2 \cdot N_{\rm e}
\; \; \; \mbox{ erg s}^{-1}
\label{power_el.eq}
\end{equation}
which should be significantly smaller than
the power supplied by the cooling flow.
The maximum possible power $P_{\rm CF}$ which can be extracted from the 
cooling flow itself can be estimated assuming a standard cooling
flow model and it corresponds to the $p \, dV/dt$ work done on the gas 
per unit time as it enters $r_{\rm c}$ : 
$P_{\rm CF} = p_{\rm c} \cdot 4 \pi r_{\rm c}^2 \cdot v_{\rm F,c}
\sim  2/5 L_{\rm cool}$ (e.g., Fabian 1994, $L_{\rm cool}$ being
the luminosity associated with the cooling region).
For typical values of the physical parameters 
in cooling flow clusters one has:
\begin{eqnarray}
\label{pcf.eq}
P_{\rm CF} &\sim& 2.3 \times 10^{43} \left(\frac{n_{\rm c}}{10^{-3} 
\mbox{ cm}^{-3}} \right) \left(\frac{kT}{\mbox{6 keV}} \right) 
\left(\frac{r_{\rm c}}{\mbox{100 kpc}} 
\right)^2 \cdot \nonumber\\
~ & \cdot & \left(\frac{v_{\rm F,c}}{\mbox{10 km s}^{-1}} \right) \; \; \;
\mbox{ erg s}^{-1}
\label{power_cf}
\end{eqnarray}
In the case of A2626, one obtains that the power necessary to 
re--accelerate the relic electrons is  
$P_{\rm e} \sim 8.3 \times 10^{40}$ erg s$^{-1}$, 
while the power supplied by the cooling flow is 
$P_{\rm CF} \sim 1.2 \times 10^{43}$ erg s$^{-1}$.
Therefore,
only a small fraction ($\sim 0.7$ \%) of $P_{\rm CF}$
should be converted into electron re--acceleration,
and thus we conclude that processes powered by 
the cooling flow itself are able to provide sufficient energy 
to power the radio mini--halo in A2626.  
A similar result is found for Perseus
(see column 10 of Table \ref{confr_ris.tab}).


\section{Observational properties of mini--halos}
\label{minihalos.sec}

\begin{figure}
\includegraphics{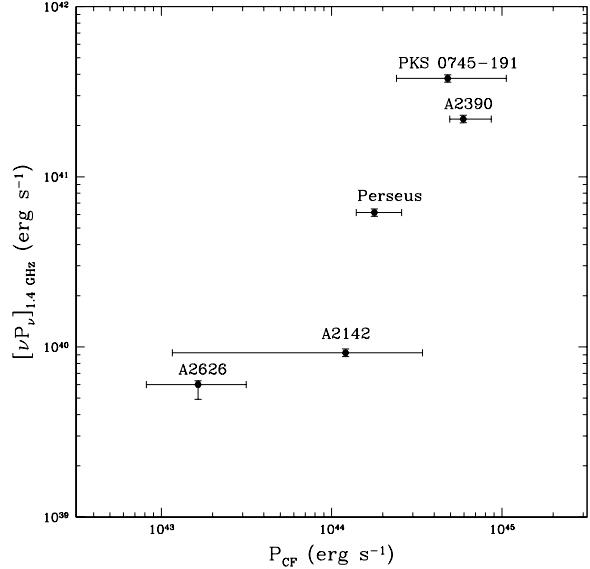}
\vspace{8cm}
\caption{
Integrated radio power at 1.4 GHz, $\left[ \nu P_{\nu} \right]_{\rm 1.4 GHz}$,
vs. cooling flow power, $P_{\rm CF} = \dot{M} k T/\mu m_{\rm p}$, 
for the mini--halo clusters in Tab. \ref{minialoni.tab}.}
\label{correlazione.fig}
\end{figure}

With the aim to explore the properties of mini--halos we selected
a small sample of candidates among known diffuse radio sources in the
literature.
The clusters in the sample were selected based on the presence of both a
cooling flow and a diffuse radio emission with no direct association
with the central radio source. 
In particular, since GBS's model assumes a connection between the origin of 
the synchrotron emission and the cooling flow, to be conservative 
we selected those clusters where the size of the diffuse radio emission
is comparable to the cooling radius. 
For this reason we excluded Abell 2052, the Virgo cluster and
2A 0335+096, which host amorphous radio sources 
with a size considerably smaller than the cooling flow region.
Relevant X--ray and radio data are reported in Tab. \ref{minialoni.tab}, 
along with references,
while in Fig. \ref{correlazione.fig} we 
report the radio power at 1.4 GHz of the mini--halos 
(in terms of integrated radio luminosity $\nu P_{\nu}$)
versus the maximum power of the cooling flows  
$P_{\rm CF} = \dot{M} k T/\mu m_{\rm p}$.
A general trend is found, with the strongest radio mini---halos 
associated with the most powerful cooling flows.

\begin{table}
\begin{center}
\caption{Observational data for mini--halos}
\begin{tabular}{lcccc}
\hline
\hline
Cluster & $z$ & $\dot{M}$ & $kT$ & $P_{1.4}$ \\
~&~&  ($\mbox{M}_{\odot} \mbox{ yr}^{-1}$) & (keV) & (W Hz$^{-1}$)\\  
\hline
~&~&~&~\\
PKS 0745-191 & 0.1028 & $579^{+399}_{-215}$ & $8.3^{+0.5}_{-5.8}$ & $2.7 \times 10^{25}$\\
Abell 2390 & 0.2320 & $625^{+75 }_{-75 }$ & $9.5^{+1.3}_{-3.4}$ & $1.5 \times 10^{25}$\\
Perseus & 0.0183 & $283^{+14 }_{-12}$ & $6.3^{+1.5}_{-2.3}$ & $4.4 \times 10^{24}$\\
Abell 2142 & 0.0890 & $106^{+248}_{-106}$ & $11.4^{+0.8}_{-3.2}$ & $6.6 \times 10^{23}$\\
Abell 2626 & 0.0604  & $ 53^{+36 }_{-30}$ & $3.1^{+0.5}_{-1.0}$ & $4.3 \times 10^{23}$\\
\hline
\label{minialoni.tab}
\end{tabular}
\end{center}
\textit{References}:
{\bf X--ray data:} 
Perseus, A2142, A2626: White, Jones \& Forman (1997) with {\it Einstein} IPC;
PKS 0745-191: White, Jones \& Forman (1997) with {\it Einstein} HRI; 
A2390: $\dot{M}$ from B\"ohringer et al. (1998) with {\it ROSAT} PSPC, 
$kT$ from White, Jones \& Forman (1997) with {\it Einstein} IPC.
{\bf Radio data:} 
PKS 0745-191: Baum \& O'Dea (1991);
A2390: Bacchi et al. (2003); Perseus: Sijbring (1993);
A2142: Giovannini \& Feretti (2000); A2626: this work.
\end{table}

Since for A2626 the only available X-ray
observation is from
{\it Einstein} IPC, 
for consistency the X--ray data were taken, when possible, from the 
compilation of White, Jones \& Forman (1997) with {\it Einstein 
Observatory}. 
For A2390, not detected as a cooling flow cluster by 
{\it Einstein} IPC, the value of $\dot{M}$ is taken from more recent 
observations with {\it ROSAT} PSPC, which have shown the presence of a 
cooling flow in this cluster (B\"ohringer et al. 1998).
Note that when both measurements from {\it Einstein} and {\it ROSAT}
are available (e.g., Perseus, A2142, PKS 0745 -191), 
the {\it Einstein}--based $\dot{M}$ is a factor $\sim$ 2 below the 
{\it ROSAT}--based value. 

We notice that the maximum powers which can be extracted from
cooling flows are orders of magnitude 
larger than the integrated radio powers (see Fig. \ref{correlazione.fig}), 
in qualitative agreement with the very low efficiencies 
calculated in the model (see column 10 of Table \ref{confr_ris.tab}).

As already discussed in Sect. \ref{reacceleration.sec},
it is worth noticing that new {\it Chandra} and {\it XMM--Newton} 
results obtained for a limited number of objects hint to an 
overestimate of $\dot{M}$ derived by earlier observations: in particular, 
the consensus reached in these studies (e.g., McNamara et al. 2000; 
Peterson et al. 2001; David et al. 2001; Tamura et al. 2001;
Molendi \& Pizzolato 2001; B\"ohringer et al. 2001; Peterson et al. 2003) is
that the spectroscopically--derived cooling rates are a factor $\sim$ 5-10
less than earlier {\it ROSAT} and {\it Einstein} values
(e.g., Fabian \& Allen 2003). 
This factor seems to be similar for all clusters in a large range of 
$\dot{M}$, giving a systematic effect that will not spoil the correlation 
reported in Fig. \ref{correlazione.fig}. 

In addition, we stress that the trend seen 
in Fig. \ref{correlazione.fig} is 
expected in the framework of GBS's model.
Qualitatively, $P_{\rm radio} \propto \langle \gamma_{\rm b} \rangle^2 \, 
{\mathcal{N}}_{\gamma_{\rm b}} \, \langle B \rangle^2 \, r_{\rm c}^3$, 
where $\langle \gamma_{\rm b} \rangle$ and 
$\langle B \rangle$ are the average values of gamma break
and magnetic field in the cooling flow region, while
$\mathcal{N}_{\gamma_{\rm b}}$ is the number density
of relativistic electrons with Lorentz factor $\gamma = \gamma_{\rm b}$.
We remind that the bulk of the observed radio emission is indeed
produced by the electrons with Lorentz factor $\gamma \sim \gamma_{\rm b}$.
On the other hand, the maximum power which can be
extracted from a cooling flow estimated on the basis of a
standard cooling flow model is
$
P_{\rm CF} = p_{\rm c} \cdot 4 \pi r_{\rm c}^2 \cdot v_{\rm F,c}
$
(see Sect. \ref{discussion.sec}), where 
$v_{\rm F,c} \propto \dot{M} \, r_{\rm c}^{-2} \, n_{\rm c}^{-1}$
(Fabian, Nulsen \& Canizares 1984). Thus from Eq. \ref{ennec.eq} one 
has: 
$
P_{\rm CF} \propto p_{\rm c} \, r_{\rm c}^3
$
and, since cooling flows are pressure--constant processes, it results:
$P_{\rm CF} \propto r_{\rm c}^3$, i.e.
$P_{\rm radio}$ is expected to increase with $P_{\rm CF}$, 
with an efficiency $\epsilon$ which depends on details (related to the 
micro--physics of the complicated parameterization of $P_{\rm radio}$) 
not considered in the model. 

The trend presented in Fig.~\ref{correlazione.fig} is based 
on few objects 
with still relatively large errors on the parameters.
If true, this trend would clearly indicate a 
connection between the thermal ICM and the relativistic electrons 
in cooling cores in qualitative agreement with our theoretical 
expectations.
It should be stressed that 
we may have introduced a bias in our sample
since, in order to deal with objects belonging to the
radio mini--halo class, we have selected only those
objects with an extension similar
to that of the cooling flow region.
These are well developed radio mini--halos which
would have a relatively high efficiency in the particle
acceleration process.
The trend between  
radio and cooling flow powers in Fig.~\ref{correlazione.fig}
may thus result from the fact that the 
efficiency of the particle acceleration 
is similar in the selected clusters.

In general, the 
efficiency in converting the cooling flow power into particle
acceleration depends on relatively unknown quantities : 
the energy transport from large--scale 
turbulence towards the smaller scales, and the details of the 
coupling between the turbulence at small scales
and the relativistic electrons.
All these quantities depend on microphysics and it may likely
be that this would lead to a situation of broad ranges of
efficiencies.

If low efficiency radio mini--halos exist, they will fill the 
bottom--right corner of Fig.~\ref{correlazione.fig}. 
These objects may be less extended and fainter than typical
mini--halos.
In addition, it may likely be that  
the electrons in these objects are  
not re--accelerated to the energies necessary to produce 
$\sim$ GHz synchrotron emission, 
and thus that they would emit only at 
much lower frequencies.
It is evident that future surveys of radio mini--halos
in cooling flow clusters combined with X--ray studies
of the ICM would shed new light on the link between 
thermal and relativistic plasma in clusters and on the
physics of turbulence and particle acceleration 
in these regions.


\section{Conclusions}

We have reported a detailed study of the radio properties
of a new mini--halo candidate in A2626.
We have shown that a particle re--acceleration model
(GBS's model)
with a set of physically--meaningful values of the parameters
is able to account for the observed 
brightness profile, the
integrated synchrotron spectrum and
the radial spectral steepening.
We conclude that A2626 has physical properties in
between the case of M87, for which there is only marginal
evidence for electron re--acceleration, 
and the prototype of mini--halos in the Perseus cluster.
Moreover, we obtain that the maximum power of the cooling flow
is more than 2 orders of magnitude
larger than the emitted radio power, thus indicating that
the cooling flow power (even if considerably reduced
by the recent observational claims) may play a leading role 
in powering the radio mini--halo.

We have selected a small sample of well developed radio
mini--halos and shown that the radio power of these objects
correlates to the cooling flow power.
GBS's model for particle re--acceleration in cooling flow
is consistent with the observed trend.
If confirmed, in the re--acceleration scenario
this trend would indicate that the conversion
of the cooling flow power into magneto--plasma 
turbulence and particle acceleration is similar 
in well developed radio mini--halos.

\begin{acknowledgements}

We thank the referee Torsten En{\ss}lin for helpful comments.
M.G. would like to thank Simone Dall'Osso for useful discussions.
M.G. and G.B. acknowledge partial support from CNR grant CNRG00CF0A.
This work was partly supported by the Italian Ministry for University
and Research (MIUR) under grant Cofin 2001-02-8773 and by
the Austrian Science Foundation FWF under grant P15868.

\end{acknowledgements}

\end{document}